\documentclass[prl,footinbib,showpacs,twocolumn]{revtex4}

\usepackage{amssymb} 
\usepackage{graphicx} 
\usepackage{dcolumn}
\usepackage{bm} 
\begin{document}

\title{Rhythmic cluster generation in strongly driven colloidal
dispersions}  \author{H. H. Wensink} \author{H. L\"{o}wen}
\affiliation{Institut f\"{u}r Theoretische Physik, Heinrich-Heine-Universit\"{a}t D\"{u}sseldorf, 
Universit\"{a}tsstra{\ss}e 1, D-40225,
D\"{u}sseldorf, Germany} \pacs{82.70.Dd, 61.20.Ja, 61.30.-v, 05.40.-a}
\date{\today}

\begin{abstract}
We study the response of a nematic colloidal dispersion of
rods to a driven probe particle which is dragged with high speed
through the dispersion perpendicular to the nematic director. In front
of the dragged particle, clusters of rods are generated which
rhythmically grow and dissolve by rotational motion. We find evidence
for a mesoscopic cluster-cluster correlation length,  {\em
independent} of the imposed drag speed. Our results are based on
non-equilibrium Brownian dynamics computer simulations and in line
with a dynamical scaling theory.
\end{abstract}

\maketitle

In recent years particle tracking techniques have been developed to
determine the viscoelasticity of soft materials on a microscale
\cite{gislerweitz-furst-review}. Optical or magnetic tweezers can be
used to drag a bead particle through the material in a controlled
way. The system's response to the tracer particle  reveals its
microrheology and provides insight into the viscoelastic  response of
biological matter  \cite{feneberg-meiners}, colloidal dispersions
\cite{habdasweeks,meyerfurst}, polymer solutions \cite{xupalmer} and
wormlike micelles \cite{jayaraman}.

Microrheological measurements are usually based on linear response,
employing the intrinsic thermal motion of the probe particle.  Much
less is known about nonlinear effects that occur  when soft materials
are subject to strongly driven probe  colloids as expressed by a high
Peclet number. The best studied examples are colloidal suspensions
thanks to their model character and direct accessibility by real-space
techniques \cite{grier-video}.  Striking novel effects in strongly
driven colloidal dispersions include  force-thinning of the
microviscosity \cite{carpenbrady},  local jamming of particles in
front of the dragged particle \cite{reichhardt-jam}, and  rhythmic
bursting of colloids driven through channels \cite{yoshi-bursting}.

In this Letter we propose a new  effect of {\em colloid cluster generation}
by  a spherical probe particle which is dragged through the suspension
with a high velocity.  In contrast to static colloid clustering which
was recently observed under equilibrium conditions
\cite{stradner-cluster,sciortino-cluster}, the cluster generation is
rhythmic here, i.e. particle clusters grow and dissolve with a
characteristic frequency.  Correspondingly, along the trace of the
moving probe particle, a characteristic cluster-cluster
correlation length $\lambda_0$ occurs.  
While the average cluster size
increases monotonically as a function of the drag velocity, 
the correlation length is found to be
{\em independent} of the velocity for strong drags. 

Our results are
obtained by Brownian dynamics computer simulations of a
two-dimensional (2D) suspension of rod-like colloidal solute
particles. The rods are in the nematic phase and  the probe particle
is dragged perpendicular to the nematic director and cannot  penetrate
the rods. We observe the following cluster generation
process at high drag speeds: the moving probe particle will touch a
rod which is then taken up to its  speed forming a  particle-rod
complex. This complex successively sweeps together further rods such
that a cluster grows in front of the dragged particle. Forcing a
collective rotation of the whole  cluster, the tracer particle is
rushing over the cluster, the cluster slowly dissolves and the process starts
again. In order to quantify and predict the statistics of this process
of dynamical cluster generation, we propose a simple dynamical scaling
theory. The simulation results for the cluster size
distribution, the time-averaged cluster size and the characteristic
cluster correlation length $\lambda_0$ are in agreement with the theory.

The rhythmic cluster generation effect is very general. It is also
expected in three dimensions and for tracer and colloidal particles of
different shapes.  The only condition is that the  interaction between
the particles possesses an excluded volume region, i.e. a region of
impenetrability. Therefore, rhythmic clustering should be observable in
many different experimental set-ups.  However, cluster generation is
absent for a soft spherical tracer particle in a suspension of soft
colloids as simulated in Ref. \cite{reichhardt-crossover}.

An important implication of the cluster generation occurs for
suspensions which coagulate once their distance falls below a critical
value. Since the average size of the clusters can be tuned by the drag
speed of the sphere, it is possible to generate aggregates with
tunable size. These may serve as building blocks for
nano-composite materials with novel rheological or
optical properties \cite{rodcluster}.

In our Brownian dynamics simulations, we consider a 2D system of $N$
charged rodlike  particles kept at constant temperature $T$ via the
solvent.  The simulation involves a numerical solution of the
overdamped  Langevin equations using the (short-time) rotational
diffusion constant $D_{0}^{r}$ and translational ones
$D_{0}^{\parallel}$ and $D_{0}^{\perp}$, parallel and  perpendicular
to the rod axis, for a single non-interacting rod with length
$L_{0}$ and thickness $d_{0}$ \cite{kirch-bd}.  The mutual rod interaction is
represented by a Yukawa-segment model in which the total rod charge is
equally partitioned over $n=13$ equidistant segments located on the
rod axis. The interaction potential between two segments $a$ and $b$
on  rods $i$ and $j$ ($j\neq i$) is of the Yukawa type, $ U = U_{0}
\exp(-\kappa r_{ab}^{ij})/ \kappa r_{ab}^{ij}$ with $\kappa$ the Debye
screening constant and $r_{ab}^{ij}$ the segment-segment
distance. The amplitude is $U_{0}=6.5 \times 10^{-2}k_{B}T$ (with
$k_{B}T$  the thermal energy) comparable to that of Tobacco Mosaic
Virus  particles in a salt-free solution under standard conditions for
which $L_{0}=300$ nm, $d_{0}=18$ nm and $\kappa^{-1}\approx 50$ nm
\cite{fraden-tmv-baus}.

A starting configuration was generated by randomly putting $N=900$
rods in a parallel, non-overlapping configuration  such that the
nematic director is parallel to the $y$-axis of a rectangular
simulation box of lengths $L_{x}$ and $L_{y}$. In all simulations the
reduced number density of the nematic system was fixed at
$\rho^{\ast}=NL_{0}^2/L_{x}L_{y} =3.0$. Equilibration typically
required $10^5$ Brownian timesteps of magnitude $\Delta t = 0.002
\tau_{0}^{T}$, measured in units of the typical translational Brownian
relaxation time of a free rod $\tau_{0}^{T}= L_{0}^{2}/D_{0}^{T}$,
with $D_{0}^{T}=(D_{0}^{\parallel}+2D_{0}^{\perp})/3$. We confirmed
that this system is in a two-dimensional nematic phase
\cite{frenkel-eppenga}  with a director along the $y$-axis of the
simulation box.

The next step in our simulation is to drag a small sphere through the
nematic host structure along the $x$-axis of the frame.  The probe
particle is a small repulsive sphere interacting with the rod segments
via a truncated and shifted Lennard-Jones potential (setting the
Lennard-Jones parameters to $\sigma_{\rm LJ}=\kappa^{-1}$ and
$\varepsilon_{\rm LJ}=k_{B}T$) and a cut-off distance
$2^{1/6}\sigma_{\rm LJ}$. The particle is trapped in a parabolic
potential well which is  displaced through the system along ${\bf \hat x}$ 
at a constant speed $v$. The  external potential for the sphere located 
at position ${\bf r}$ can be written as
$U_{\rm{ext}} ({\bf r},t)= A_{0}|{\bf r} - vt {\bf \hat x}|^{2}$
with a large amplitude $A_{0} = 10^{6}k_{B}T/L_{0}^2$ for the trapping
potential.  The particle is dragged until it has traversed the
horizontal box size $L_{x}$. Henceforth, we will use a reduced drag
speed $v^{\ast}=vL_{0}/D_{0}^{T}$ which expresses $v$ in units of the
Brownian speed of a freely diffusing rod. Quantitative results were
collected by averaging over consecutive independent  initial
configurations for the  nematic state.
\begin{figure}
\includegraphics[width=8.5cm,height=7.5cm]{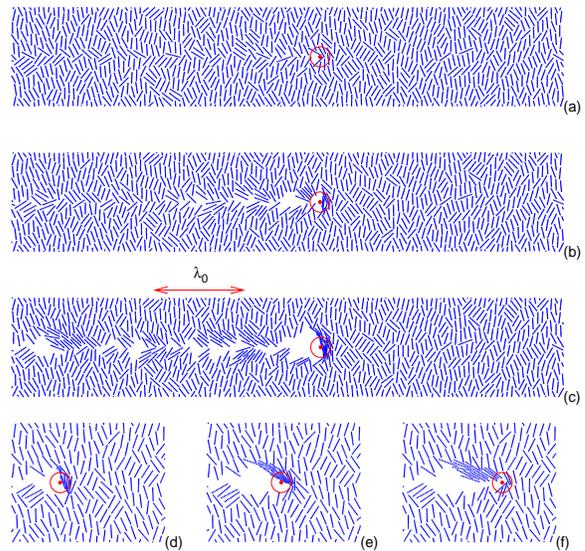}
\caption{ Snapshots of the nematic structure at  (a) $v^{\ast}=100$, (b)
$v^{\ast}=1000$ and  (c) $v^{\ast}=4000$. The driven sphere is
indicated by the encircled dot. The sequence (d)-(f) depicts the growth 
and rotation of a single cluster at $v^{\ast}=4000$. The length of the arrow 
corresponds to the cluster-cluster correlation length of $\lambda_{0}=5L_{0}$. }
\end{figure}

In Figs. 1a-c  three simulation snapshots are depicted corresponding to 
a dragging event at increasing drag speeds. At high drag speeds a  rhythmic
clustering effect is observed, as reflected in Figs. 1b and c, in which rods 
are pushed together into
an anisometric cluster which subsequently dissolves by collective
rotation under the influence of the force exerted by the dragged
sphere. This process is depicted explicitly in a time series of snapshots shown in Figs. 1d-f.
In all snapshots, voids can be observed in the wake of the dragged particle which slowly dissolve over time.
The length  of the void trail  grows continuously with drag speed. The same holds for the void size indicating that the time
needed for the system to relax back to equilibrium also increases with drag speed.

The instantaneous cluster size can be monitored
using a simple cluster criterion based on the distance of closest
approach between rod pairs located in front of the drag sphere. A rod
pair is assigned to a cluster if this distance is smaller than 0.5
times the average distance of closest approach in the equilibrium
nematic phase. The rhythmic nature of the clustering is clearly
reflected in the cluster time series depicted in Fig.  2. Intermittent
behavior is evident from the sudden outburst of  a big clustering
event after a sequence of smaller ones. We verified that $N_{c}(vt)$
does not significantly change upon small variations of the cluster selection criterion.
Defining $\langle .. \rangle$ as an average over time $t$ we observe that $\langle N_{c}(vt) \rangle$ increases smoothly as a 
function of the drag speed implying that the transition towards a state of rhythmic clustering occurs continuously. 
\begin{figure}
\includegraphics[width=7.5cm,height=4.5cm]{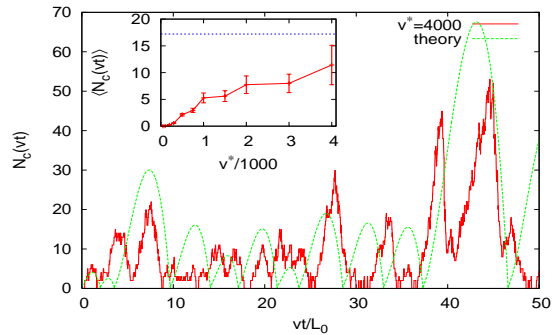}
\caption{Time series for the cluster size $N_{c}$ for a single drag
event. Inset: time averaged cluster size $\langle N_{c}(vt)\rangle$ versus drag speed. The dashed
line $\langle N_{c}\rangle =17.2$ follows from theory.}
\end{figure}
To investigate correlations between the cluster sizes we have
carried out a spectral analysis of the cluster time series.
The power spectrum $S(q)$ represents the Fourier transform of the cluster-cluster autocorrelation function
via $S(q)=\int_{-\infty}^{\infty}dt'\exp[-iqvt']\langle N_{c}(vt)N_{c}(v(t+t'))\rangle $ with $q$ the wavenumber.
\begin{figure}
\includegraphics[width=8.5cm,height=4.2cm]{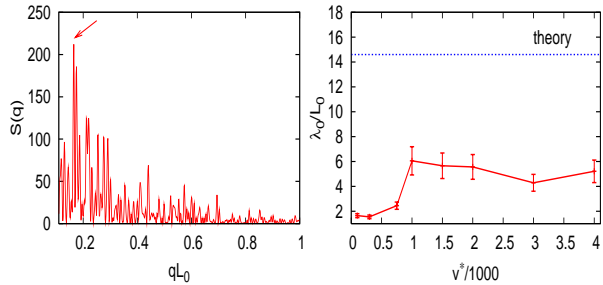}
\caption{(a, left) Power spectrum for $v^{\ast} =1000 $. The correlation peak is indicated by the arrow.  (b, right) Dominant
cluster-cluster correlation length $\lambda_0$  corresponding to the
maximum amplitude of the power spectrum versus  speed. The horizontal
line is the theoretical result $\lambda _{0} \approx 14.6 L_0 $.}
\end{figure}
In the relevant spectral range, all spectra show a dominant peak (as indicated in Fig. 3a) which we can identify
 with a typical cluster-cluster {\em correlation length} 
 $\lambda_{0}$.  Whereas the dominant peak is blurred somewhat by the
 high noise level in the spectra at low speeds, it becomes much more
 pronounced at $v^{\ast}>1000$, see Fig. 3a. The location of the
 dominant peak for various $v$ is presented in Fig. 3b.   A striking effect is that
 $\lambda_{0}$ does not increase with speed but remains at constant
 value of  roughly $5-6L_{0}$.  This result suggests the existence of
 a typical cluster-cluster correlation length scale for the nonlinear response
 of a driven system of colloidal rods. The length scale is also indicated in Fig. 1c.

 Given the generic character of
 the clustering mechanism, we expect this length scale to be
 insensitive to the details of rod pair interactions and to depend
 only on the concentration $\rho^{\ast} $of the nematic phase \cite{note}.
Moreover, we anticipate dynamic clustering to be {\em unique} for
rodlike systems and do not expect it to occur in dense systems of soft
spheres. The reason for this is that two charged rods in 2D have a
finite core excluded volume in the limit $\kappa^{-1}\searrow 0$,
namely $V_{\rm ex}=2L_{0}^{2}\sin \gamma$ (with $\gamma$  the angle
between the rods), unlike soft Yukawa spheres for which $V_{\rm
ex}=0$. A finite exclusion zone around each particle is a necessary
condition for the formation of large force chains that are associated
with the clustered particles.  To verify this statement we have
performed similar drag experiments for dense fluids of Yukawa spheres
close to the glass transition for various Lennard-Jones parameters for
the sphere-rod interactions.  In all cases, no evidence for rhythmic
clustering was found up to drag speeds exceeding the present range by
several orders of magnitude.

We now propose a simple dynamical scaling theory for the observed
behavior. At high drag speeds the external forces due to the dragged
spheres are much larger than the Brownian forces and the system is
governed by overdamped, deterministic motion. The rotational motion of
a {\em single} rod-like cluster with length $\ell$ under an external
torque $\cal T$ is then given by $\xi_{r}(\ell){\dot \varphi(t)}=
{\cal T}$ with ${\dot \varphi}$ the time derivative of the angle
between the rod axis and ${\bf \hat y}$ and $\xi_{r} = \ell^3{\cal F}(p)$
the rotational friction coefficient of the cluster, with  ${\cal F}$ a
function of the {\em cluster}  aspect-ratio $p$ \cite{tirado}.
\begin{figure}
\includegraphics[width=8cm,height=4cm]{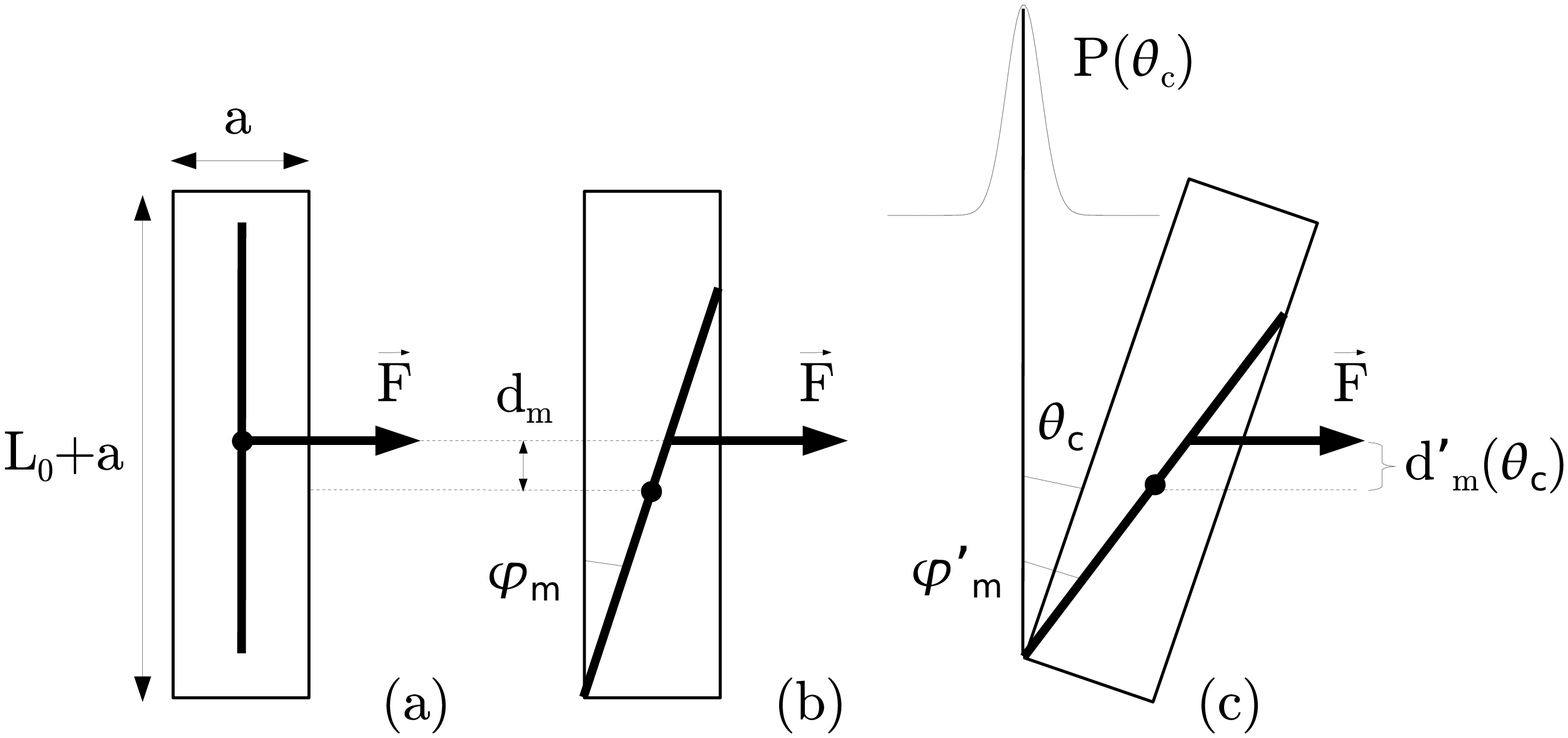}
\caption{Sketch of the cage model used  in the  scaling theory.}
\end{figure}
 The force $F$ directed along the $x$-axis is proportional to $\ell
 \cos \varphi$ and the velocity, so we write  $F= v\xi_{\perp}^{0}\ell
 \cos \varphi$, with  $\xi_{\perp}^{0}$ the translational friction
 coefficient of an individual rod. The torque is then given by the
 force times the vertical component of the arm length $d$. We consider
 $d$ to be a {\em dynamical} angle-dependent variable proportional to
 speed, i.e.  ${\dot d}=v \sin \varphi$. Finally, we assume the growth
 of the cluster length to be   proportional to the speed while an
 increased inhibition is expected as the cluster rotates towards the
 $x$-direction. This gives  ${\dot \ell}=vL_{0}a^{-1} \cos
 \varphi$, where the additional length scale $a$ pertains to the nearest neighbor
 distance in the nematic phase. It can be quantified from a simple cell description as sketched in Fig. 4a.
If we assume each rod to be confined to a rectangular cage, we can directly obtain $a^{\ast}=a/L_{0}$ from the
concentration of the nematic phase, via
$2a^{\ast}=(1+4/\rho^{\ast})^{1/2}-1$, giving $a^{\ast}=0.264$ for
$\rho^{\ast}=3$. We also assume the cluster aspect-ratio $p$ to be a constant
inversely proportional to $a^{\ast}$, i.e. $p=1/a^{\ast}$. 
Gathering all contributions gives a set of three coupled dynamical nonlinear equations:
\begin{eqnarray}
v^{-1}{\dot \varphi } &&= \Theta (x) \xi_{\perp}^{0} {\cal F}^{-1}  \ell^{-2} d \cos \varphi
\nonumber \\ v^{-1}{\dot d } &&= \sin \varphi \nonumber  \\
v^{-1}{\dot \ell} &&= {a^{\ast}}^{-1} \cos \varphi \label{model}
\end{eqnarray}
 with initial values $\{ \varphi_{0},d_{0},L_{0} \} $ at $t=0$. It is
important to note that the velocity can be {\em scaled out} of
Eq. (\ref{model}) using the total drag distance $vt$ as the
independent variable. A stepfunction $\Theta(x)$ with $x=(\ell /2) \cos \varphi-d$ has been
included for consistency reasons to enforce the rotation and growth
process to stop once the arm length exceeds $\ell/2$. 

Due to the external force the rod is pushed
toward the edges of the cell, indicated in Fig. 4b,  and we may define
a {\em typical} cage deflection angle given by $\varphi_{\rm m}=\sin
^{-1} a^{\ast}$ and  associated arm component $d^{\ast}_{\rm
m}=(a^{\ast}+1-\cos\varphi_{\rm m})/2$.   If we use these as the
initial values for Eq. (\ref{model}) we numerically obtain a dominant
cluster length scale of $\lambda_{0}=vt({\dot \varphi}=0)=14.6L_{0}$
in qualitative agreement with the simulation data, see Fig. 3b. The 
overestimation is probably due to the somewhat crude nature of the
cage model and the neglect of Brownian fluctuations which are expected
to play a role in the initial stages of cluster formation.

Fluctuations around the dominant length scale can be introduced by
noting that the rods are not all parallel  due to local orientational
fluctuations around the nematic director ${\bf \hat y}$. Therefore, we can
introduce an angle $\theta_{c}$ measuring the deflection angle of the
cage from its parallel orientation $\theta_{c}=0$.  The fluctuation
strength can be roughly quantified by a nematic order parameter $0\leq
S_{z} \leq 1$, defined as $S_{z}=\langle \sum_{i} \cos 2 \theta _{i}/N
\rangle$ with $0 \leq \theta_{i} \leq \pi/2$ the angle between rod $i$
and ${\bf \hat y}$. We measured $S_{z}\approx 0.8$.  Assuming the cell
orientation angle  $\theta_{c}$ to obey a  Gaussian distribution
$P(\theta_c)\propto \exp[-\theta_{c}^{2}/2(1-S_{z})^2]$, we can
construct a series of cluster events by solving Eq. (\ref{model}) for
$N=1000$ consecutive sets of initial values $\{\varphi_{\rm
m}+\theta_{c}, d_{\rm m}^{\prime}(\theta_c),L_{0}\}$, with $0 \leq
\theta_{c} \leq \pi/2$ sampled {\em randomly} from the Gaussian
distribution (see also Fig. 4c).  The cluster size $N_{c}$ can be 
obtained by multiplying the area fraction of the
cluster rectangle located ahead of the drag particle  with the typical rod
concentration $\rho^{\ast}_{c}$ inside the cluster,  which we estimate
at $\rho^{\ast}_{c} \approx 6.0 $ from our simulations.  From the time
series (included in Fig. 2) we can extract the average cluster size 
$\langle N_{c} \rangle$ (Fig. 2) and the distribution function of the 
cluster size shown in Fig. 5. It can be deduced from Fig. 5 that the strong tail 
of the  distributions at large speeds is correctly reproduced by the theory.
\begin{figure}
\includegraphics[width=7.5cm,height=4.5cm]{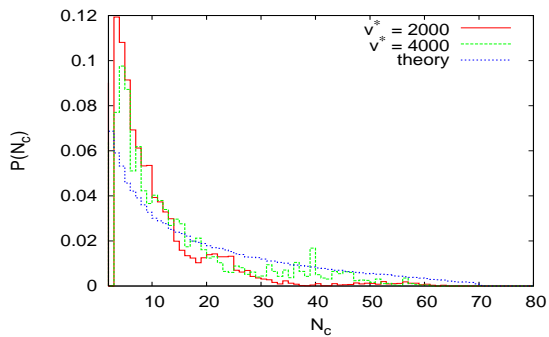}
\caption{Distribution of the cluster size $N_{c}$ for large speeds.}
\end{figure}

In summary, we have studied the effect of a small colloidal sphere
driven through a nematic background at various drag
speeds. We observe the formation  of rod clusters whose growth is
inhibited by collective rotation of the cluster under the influence of
the force exerted by the dragged sphere.   The growth-inhibition
mechanism gives rise to {\em rhythmic} clustering behavior, quantified
by a typical cluster-cluster correlation length which is independent
of the imposed drag speed. A simple dynamical model without adjustable
parameters is proposed based on overdamped rotational motion of a
single rod cluster. The theory predicts a velocity-independent scaling
value for the cluster-cluster correlation length in qualitative
agreement with the simulation results.

The observations described in this Letter can be
verified in experiments.  Various experimental realizations are
conceivable: one can confine suspensions of rod-like particles between
two parallel glass plates and use an optical tweezer to drag a probe
particle  through the suspensions. Confocal microscopy and
video-microscopy can be used to follow cluster generation dynamically.
Further set-ups to test our  predictions are colloidal nematic
monolayers on substrates \cite{silvestre}, nematics anchored at
interfaces \cite{jerome}, quasi-2D granular rods \cite{granularrods}
or elongated dust rods levitated in a  plasma sheath
\cite{dustyplasma}. Qualitatively similar rhythmic clustering is expected 
in 3D colloidal nematic systems if a thin transversely
oriented {\em rod} is driven perpendicular to the nematic director.

We thank E. Weeks, E. Kaler, E. Furst and K. Yoshikawa for fruitful
discussions. HHW acknowledges the  {\em Alexander von Humboldt Foundation}
for financial support.  This work is part of the SFB-TR6 (project D1).


\begin{thebibliography}{19}
\bibitem{note} At constant $\rho^{\ast}$, the cluster size does not depend strongly on
the rod aspect ratio $L_{0}/d_{0}$ as long as the system is in the nematic state.
\end{thebibliography}
\end{document}